CellPhoneDB v5: inferring cell-cell communication from single-cell multiomics data

## Authors

Kevin Troulé 1, Robert Petryszak 1, Martin Prete 1, James Cranley 1, Alicia Harasty 2, Zewen Kelvin Tuong 1,2, Sarah A Teichmann 1,3, Luz Garcia-Alonso 1,#, Roser Vento-Tormo 1,#

1 Wellcome Sanger Institute, Cambridge, UK
2 Ian Frazer Centre for Children's Immunotherapy Research, Child Health Research Centre, Faculty of Medicine, The University of Queensland, Brisbane, QLD Australia
3 Department of Physics, Cavendish Laboratory, University of Cambridge, JJ Thomson Ave, Cambridge

# co-last and co-corresponding (lg18@sanger.ac.uk, rv4@sanger.ac.uk)

## Abstract

Cell-cell communication is essential for tissue development, regeneration and function, and its disruption can lead to diseases and developmental abnormalities. The revolution of single-cell genomics technologies offers unprecedented insights into cellular identities, opening new avenues to resolve the intricate cellular interactions present in tissue niches. CellPhoneDB is a bioinformatics toolkit designed to infer cell-cell communication by combining a curated repository of *bona fide* ligand-receptor interactions with a set of computational and statistical methods to integrate them with single-cell genomics data. Importantly, CellPhoneDB captures the multimeric nature of molecular complexes, thus representing cell-cell communication biology faithfully. Here we present CellPhoneDB v5, an updated version of the tool, which offers several new features. Firstly, the repository has been expanded by one-third with the addition of new interactions. These encompass interactions mediated by non-protein ligands such as endocrine hormones and GPCR ligands. Secondly, it includes a differentially expression-based methodology for more tailored interaction queries. Thirdly, it incorporates novel computational methods to prioritise specific cell-cell interactions, leveraging other single-cell modalities, such as spatial information or TF activities (i.e. CellSign module). Finally, we provide CellPhoneDBViz, a module to interactively visualise and share results amongst users. Altogether, CellPhoneDB v5 elevates the precision of cell-cell communication inference, ushering in new perspectives to comprehend tissue biology in both healthy and pathological states.

Introduction

Modelling cell-cell communication (paracrine, juxtacrine and endocrine) from single-cell genomics atlases provides a unique opportunity to get holistic views of tissue physiology *in vivo*. To reconstruct cellular crosstalk from single-cell transcriptomic data, we developed CellPhoneDB[1,2], which estimates the abundance of ligands and receptors from single-cell gene expression. For this, our tool combined three fundamental parts. First, a repository of reliable molecular interactions with known roles in cell-cell communication. Second, computational methods to integrate the repository with transcriptomic data to infer tissue-specific or context-specific cell-cell interactions. Third, approaches to interrogate, prioritise and visualise the results, which are often overwhelmingly dense and challenging to interpret. Using CellPhoneDB, researchers have been able to provide clues to understand the etiopathology of human diseases[3–6] as well as to use it as a guide to design therapeutic strategies[7,8] and improve *in vitro* models[9–11]. Here we present an updated version of CellPhoneDB v5, which significantly improves both the database as well as the computational methods to infer, prioritise and visualise cellular interactions.

Key for cell-cell communication inference is the accuracy and the scope of the prior knowledge covered. Inaccurate or incomplete collection of molecular interactions will negatively impact the interaction inference, independently of the tool or method used. CellPhoneDB was the first tool to infer cell-cell communication tool from single-cell genomics that contained a repository of *bona-fide* manually curated interactions from the scientific literature and peer-reviewed repositories such as Reactome[12], UniProt[13] and HMRbase[14]. During the curation process, we accounted for the subunit architecture of the proteins involved in the interaction to represent heteromeric complexes accurately. This is critical because: (i) some proteins can only work as heteromeric complexes, meaning that the interaction is not functional if one of the subunits is absent, and (ii) multi-subunit complexes change their specificity depending on the combination of heteromers so that different complex combinations bind different ligands and activate different downstream signals or TFs[15]. Thus, the inclusion of heteromeric complexes reduces the number of false positive interactions by approximating more reliably the nature of cell-cell interactions. In addition to expanding the number of interactions stored in CellPhoneDB, in this new version, we also include interactions that are mediated by small non-protein molecules by examining the last steps of their biosynthetic pathways[16–18].

There is no universal method to infer cell-cell communication from single-cell transcriptomics atlases that applies to all research scenarios. Instead, the method of choice should be tailored to the experimental design and the research question of interest. To increase the versatility of CellPhoneDB, we have implemented new computational approaches to infer cell-cell interactions based on differential expression analyses[19]. This method allows researchers to take a more tailored approach, enabling them to address their specific biological questions with greater precision. Additionally, because of the amount of cell-cell interactions is often large scale, we have included novel methods to prioritise interactions by leveraging information of other omic modalities; (i) spatial transcriptomics to restrict the analyses to those cell type pairs coexisting in time and space[19], (ii) single cell ATAC-sequencing (scATAC-seq) to prioritise interactions that are more likely to be functional based on the downstream activation status of their targeted TF (CellSign module[17]). We have also incorporated strategies to score interactions according to their expression specificity in a given cell-type pair. Finally, we have implemented a set of strategies to perform selective queries and visualisations facilitating results interpretation and sharing (CellPhoneDBViz interface[20]). Taken together, CellPhoneDB v5 offers more tools and methods to infer and prioritise cell-cell interactions, making it versatile for various experiments.

Updates on the development of the protocol: the database

As highlighted in[16–18], we have expanded our CellPhoneDB repository to include cell-cell interactions mediated by non-protein ligands that are not encoded by a gene such as steroid hormones (e.g. oestrogen), neurotransmitters (e.g. glutamate) and other small molecules (e.g. histamine). The majority of those small molecules bind G-protein-coupled receptors (GPCR) which are the largest family of membrane receptors that are targeted by approved drugs[21]. Small molecules (or metabolites) are produced by a chain of enzymatic reactions, generally coupled with a transporter, that convert the original substrate into the final metabolite[22]. This small molecule (or metabolite), in turn, will act as a ligand mediating cell interactions. To approximate this modality of cell-cell interactions and estimate the abundance of non-protein ligands, we use the last *bona fide* enzyme in the biosynthesis pathway and, if known, the transporter.

In addition, we have included more interactions by manual curation and added novel interactions coming from a recent high-throughput surface receptor screening that takes into account the affinity between proteins and their natural stoichiometry[23]. To ensure only interactions with validated roles in intercellular communication are included and minimise the number of false positive predictions, this version of the database only contains manually curated interactions and no longer collects general protein-protein interactions from third parties such as IUPHAR[24], InnateDB[25] or the IMEx Consortium[26]. This annotation strategy, combining manual curation of the interactions and considering the combinatorial specificity of protein complexes, approximates more reliably the biological nature of cell-cell interactions, reducing the number of false positive interactions inferred by the computational methods. As in previous versions, the user can modify the internal database to include new interactions facilitating more tailored queries of the dataset.

Taken together, CellPhoneDB v5 repository includes 240 heteromeric proteins, 1,058 homomeric proteins and 120 non-peptidic ligands of which 61 are composed of multiple proteins (enzymes and transporters) (Figure 1). CellPhoneDB v5 repository sums up a total of 2,911 manually curated interactions, of which 1,882 are mediated exclusively by proteins while the remaining 1,029 are mediated by non-peptidic ligands. Out of the 1,882 protein-protein interactions, 905 are mediated by heteromeric proteins, while of the 1,029 interactions mediated by non-peptidic proteins 433 are participated by heteromeric complexes.

Moreover, in this database version, we have organised all the interactions to properly reflect the interaction directionality (from sender to receiver cells) by fetching the protein location and function from the Uniprot database (UniProt Consortium 2021) followed by manual revision. Directionality is key to tracking back signalling cascades and connecting them to TF measurements. A total of 2,508 interactions have been classified as directional (ligand-receptor), 242 as adhesion interaction and the remaining 161 as either ligand-ligand, receptor-receptor or gap junction. Finally, by leveraging information from Reactome followed by manual curation we have classified the interactions into 164 functionally related pathways (only pathways with at least two interactions are kept).

## Updates on the development of the protocol: computational methods to infer and prioritise cell-cell interactions

There is no "one size fits all" manner to estimate cell-cell communication from single-cell transcriptomics data, and therefore, we offer three methods to integrate the repository with single-cell transcriptomics data (Figure 2). The user could select the method of choice depending on the experimental design and the research question of interest ). Method 1 (simple analysis) returns the mean expression of the interacting partners for each pair of cell types. For multi-subunit complexes, the method uses the expression value of the least expressed partner. Users can exclude interactions in which at least one gene member is expressed by less than a user-specified fraction of cells in a cell type. This method returns no statistics. Method 2 (statistical analysis), is an extension of the previous method that evaluates the significance of the interaction means (as calculated in method 1) by applying an empirical random shuffling approach to estimate the null distribution of the means and that was implemented in the first version of CellPhoneDB[1,2]. This is performed by randomly permuting the cluster labels of all cells in the dataset, meaning all cell types are used to estimate the null distribution. The P-value of each interaction between a cell type pair is calculated as the proportion of means in the null distribution that are equal or higher than the observed mean. To reduce the computational time on big datasets, CellPhoneDB can perform subsampling using a geometric sketching[27] of the original data, reducing the size of the original dataset while preserving rare cell states. Finally, method 3 (differential expression analysis) implemented in[19], offers an alternative approach to the statistical method 2. This new approach allows the design of more complex and flexible queries (for example, restricting the cell types in the null distribution) to identify relevant interactions for specific cell types. Thus, it can be of benefit to identify interactions (i) specific to individual cell states within a lineage[18,19], (ii) specific to cells of different conditions (i.e. disease vs control; perturbation vs steady-state) (iii) or arising during a differentiation process. It requires a user-provided list of cell types of interest and their differentially expressed genes, which can be estimated by comparing discrete cell types or along a pseudotime trajectory. Thus, differential expression analysis is pre-computed by the user using their preferred method, in a way that fits their experimental design and research question. Using this approach, CellPhoneDB will retrieve those interactions in which at least one gene-cell type pair is in the user-provided differentially expressed list and the other gene is expressed by another cell type above the user-specified threshold.  All methods rely on single-cell transcriptomics previously annotated by the user, where cells are grouped into discrete groups, which could be cell types or any functional unit defined by the user.

Inspired by  other cell-cell communication tools such as CellChat[28] or IcellNet[29], CellPhoneDB v5 includes a scoring methodology that prioritises cell-cell interactions based on their gene expression specificity. This allows the user to shorten the extensive amount of interactions generated. In essence, the scoring method of CellPhoneDB starts by calculating the mean expression of each gene per cell type, excluding those genes expressed by less than a user-specified fraction of cells within each cell type. For heteromeric proteins, CellPhoneDB aggregates the mean expression of genes participating in a complex employing the geometric mean. By doing so, the expression of complexes in which any of their respective subunits are absent is set to 0, thus excluding non-functional complexes. Following this, the mean expression of the proteins is scaled between 0 and 10 across all cell types. Those cell types with the highest mean expression for any given protein will have a scaled expression value of 10, while those with the lowest expression will have a value of 0. Finally, the product of the mean scaled protein expression between each cell pair and interacting proteins is calculated as a proxy for their interaction specificity, yielding a score ranging between 0 and 100. Interaction scores closer to 100 are indicative of higher expression specificity of the interacting proteins in a given cell pair.

To prioritise cell-cell interactions, CellPhoneDB v5 also relies on the integration of additional genomics multimodal datasets. Described for the first time in[19], CellPhoneDB v5 uses spatial transcriptomics to restrict the analyses to those cell type pairs coexisting in time and space, hence reducing the number of interactions to those occurring between neighbouring cells. Spatial microenvironments can be defined from multiple data modalities that identify colocalizing cells, including prior biological knowledge (for example, pericytes co-localize with endothelial cells in the vessels[30]), or in a data-driven manner from spatial transcriptomics (for example, 10x Visium combined with cell2location[31], stereoscope[32] among other tools[33]) and imaging-based approaches. Methods to systematically infer cellular niches whose results can be used to define spatial niches for CellPhoneDBs include, SpatialDE2[34], Giotto[35] or SPARK[36]. Temporal dynamics can also be taken into account for datasets atlasing as well as developmental processes, experimental conditions (i.e. perturbation vs control) or cohorts (i.e. disease vs control) to exclude cell types that do not coexist in time.

In addition, we have implemented the CellSign module described in[17], which identifies activated receptors and prioritises high-confidence interactions by leveraging the activity of the downstream TFs. CellSign relies on a database of receptors linked to their putative downstream TFs. These receptor-to-TF relationships are retrieved from a manual revision of the literature and are restricted to TF with high specificity for an upstream receptor. Currently, the resource contains a total of 212 highly-specific direct receptor-to-TF relationships. CellSign module uses TF activation as a downstream sensor for receptor activation upon cell-cell interaction, thus adding an extra layer of evidence to the likelihood of the cell-cell interaction. The module requires a user-provided list of the TFs that are active in each cell type, ideally estimated in a data-driven manner. Active TFs can be defined by the user by leveraging: (i) their own TF expression (ii) the expression of their target genes (DoRothEA[37]) (iii) the chromatin accessibility of their binding motifs from scATAC-seq atlases (ChromVar[38],

SCENIC[39]). The receptor-TF database can be updated by the user to include additional relationships of interest.

Updates on the development of protocol: the visualisation tool, CellPhoneDBViz & ktplots

CellPhoneDBViz is a new implementation implemented in[20] for providing web-based, interactive visualisations and search through the results of any CellPhoneDB analysis, with a minimum amount of configuration by the user. The resulting web visualiser can be used to share CellPhoneDB results with other researchers with no need of computational knowledge, and allow them to perform interactive searches, display holistic views to uncover global interaction patterns or zoom into specific cell type pairs or microenvironments to perform more specific queries. Specifically, it provides a variety of different visualisations, including heatmaps, chord plots and dot plots that can represent both CellPhone inputs and outputs. Inputs may encompass the cell composition of the full cell atlas of the spatiotemporal microenvironments. Meanwhile, outputs detail the total number of interactions between cell types, which can be split by micro-environments if data is available, and individual mean expressions across user-selected interactions and cell type pairs. The plots can be filtered by genes, interacting pairs, cell types and cell type pairs, as well as microenvironments - if applicable and available. Plots can be downloaded as pdfs.

In addition to using CellPhoneDBViz , visualisation of the results from CellPhoneDB can be achieved with ktplots (R) and ktplotspy (python)[40] packages. Each package contains functions that will accept the output from CellPhoneDB, including the data frames corresponding to interaction means, p-values (or relevant interactions), deconvoluted values, and the new output from the interaction scores and CellSign modules. Both packages also require the corresponding single-cell object used for CellPhoneDB (e.g. Seurat, SingleCellExperiment, AnnData). The primary function "plot_cpdb" plots the results as a dot plot where the size and colour gradient of each interaction for each cell-type pair is scaled according to the significance and the mean value of the interacting partners respectively. If results from the interaction scores and CellSign modules are provided, "plot_cpdb" will scale this interaction by gradient, highlighting the more relevant/significant interactions, and/or allow for the ability to filter the interactions. Other associated functions use the long-form data frame generated from "plot_cpdb" to generate chord diagrams and various other visualisations. All the plots are customisable because they are primarily created using ggplot2 (in R) and/or plotnine and matplotlib syntax. Finally, ktplots also implements the original CellPhoneDB visualisation, which showcases a heatmap of the total number of interactions per cell-type pair. This visualisation can be achieved using pheatmap (R) or seaborn (python)) and offers both symmetric and non-symmetric outputs.

Applications of the method

The updates included in CellPhoneDB v5 have been successfully used to study cell-cell communication across multiple biological contexts. By incorporating non-peptidic ligands, CellPhoneDB broadens the range of contexts it addresses, particularly in areas like endocrine and neuroendocrine biology. For instance, we used the tool to study the crosstalk between oocytes (female eggs) and developing granulosa cells (ovarian supporting cells) during fetal gonadal development[17]. Additionally, CellPhoneDB has been instrumental in identifying pacemaker cell interactions in the human sinoatrial node, revealing an unexpected synaptic interaction with niche-partner glial cells[16].

The addition of two new computational methods to infer cell-cell interactions in CellPhoneDB has facilitated the query of new datasets by allowing for more customised analyses. Method 2 has been used to identify 15 monocyte cytokine-mediated interactions specific to patients with severe COVID-19[41]. This method also helped elucidate the differential usage of ligands and receptors between two groups of non-small cell lung cancer. These groups were classified according to a lung cancer activation module (LCAM), providing deeper insights into immune cell crosstalk underlying the LCAM tumors[42]. Conversely, we have used method 3 to identify the cell-cell communication events present in the human placental-uterine interface[18]. This method pinpointed the ligand-receptors driving the differentiation of trophoblast, the placenta's epithelial-like cells. To do this, we calculated the differential gene expression between trophoblast cell states, tracing back along g their pseudotime trajectory. This analysis illuminated the signalling pathways activated during trophoblast differentiation and underscored the influence of the maternal uterine microenvironment in this process.

Introducing methods to prioritise interactions has streamlined the interpretation of our results. In the context of ovarian development[17], we utilised the spatiotemporal microenvironments modules to shed light on how the compartmentalisation of the human ovary by granulosa cell subtypes influences oocyte differentiation[17]. Similarly, when analysing placental development[18], microenvironments were proven valuable to prioritise interactions between invading trophoblast and maternal uterine cells. We also used CellSign for the prioritisation of interactions between oocytes and granulosa cells in our gonadal cell atlas[17]. Our first step was using the scATAC-seq data to pinpoint active TFs in each cell subtype.

With CellSign we then gave priority to interactions where the TFs downstream of the receptor were active.

Implementation and software availability

The database available in CellPhoneDB v5 can be accessed through a user-friendly interface at [www.cellphonedb.org](www.cellphonedb.org). Raw data can be found at [https://github.com/ventolab/CellphoneDB-data](https://github.com/ventolab/CellphoneDB-data). The complete CellPhoneDB v5 toolkit, encompassing both the database and computational methods, is available at [https://github.com/ventolab/CellphoneDB](https://github.com/ventolab/CellphoneDB) and as a PyPi package. CellPhoneDB v5 requires python version 3.8 or greater, and the latest versions of *pandas* (v1.5.0), *scanpy* (v1.9.1), *anndata* (v0.8) and geosketch (v1.2). For optimal use, CellPhoneDB v5 is tailored to run through Jupyter notebooks.

CellPhoneDBViz was developed using the D3 JavaScript library and Materialize framework, and it can be accessed from PyPi or directly from GitHub at https://github.com/datasome/cellphonedbviz/. Comprehensive documentation is included. In addition, the python and R packages *ktplotspy* (v0.3.0) and *ktplots*[43] (v3.0.0) to visualise cell-cell interactions can be accessed at [https://www.github.com/zktuong/ktplots](https://www.github.com/zktuong/ktplots) and [https://www.github.com/zktuong/ktplotspy](https://www.github.com/zktuong/ktplotspy) respectively. Detailed documentation and tutorials are available at [https://zktuong.github.io/ktplots/](https://zktuong.github.io/ktplots/) and [https://ktplotspy.readthedocs.io/](https://ktplotspy.readthedocs.io/).

CellPhoneDB v5 has been optimised for better performance, particularly in its most time-consuming statistical analysis, method 2. We have transitioned from data frame structures to numpy matrices for more efficient data handling. Additionally, the scoring method can be run in parallel, thanks to the multithreading, which accelerates the computational process Before initiating an analysis, users can also access a tool that estimates the RAM memory required for each type of analysis, based on the number of cells in the input file. Comprehensive documentation for the package is available at [https://cellphonedb.readthedocs.io/en/latest/](https://cellphonedb.readthedocs.io/en/latest/).

## Limitations

CellPhoneDB compiles an extensive list of manually curated experimentally derived cell-cell interactions with special emphasis on representing the structural composition of multimeric proteins. With its accompanying computational resources, CellPhoneDB has proven to be a robust tool for inferring cell-cell interactions[44]. It has played a pivotal role in generating new hypotheses, many of which have been experimentally validated using in vitro models[19].

Despite this success, the inference of cell-cell interactions from single-cell data remains a complex task. Firstly, cell-cell interactions are mediated by proteins and metabolites whose abundance is not measured directly. Instead, we approximate their abundance using gene expression measurements which do not necessarily account for protein translation, processing, post-transcriptional modifications (such as phosphorylation or glycosylation), secretion and diffusion. Secondly, the specificity of the signalling pathway is largely determined by the structural composition of the interacting proteins[45], and a protein can be part of different complexes with opposite effects[46]. Thirdly, cell-cell interactions are fine-tuned by various cofactor molecules, both inhibitory and activatory. The specific function of these cofactors is often not well understood, leading to ambiguities in their physiological roles and context-dependent specificity. Given these complexities, any cell-cell interaction database, no matter how comprehensive, will never capture every potential interaction, and this should be kept in mind during analysis interpretation. CellPhoneDB, specifically, relies on experimentally supported interactions. While this may lead to a bias towards well-studied processes, it is also a strength. This approach reduces the chances of false positives and ensures the results of interactions that have a role in cell-cell communication.

Inferring cell-cell communication results in high-throughput data, making visualisation and interpretation challenging. For clearer insights, it is crucial that researchers set a specific biological question beforehand, rather than using these tools for indiscriminate 'result fishing'. To facilitate this task, CellPhoneDB v5 includes novel methods to prioritise interactions based on other single-cell data modalities or prior knowledge. Additionally, a scoring methodology is included to rank cell interactions by their expression specificity.

## Comparison with other approaches

There are multiple computational methodologies to infer cell-cell interactions from single-cell datasets that emerged after CellPhoneDB[1]. Broadly, these methods fall into two categories, those that rely on a curated set of interactions, and those that do not. The curated interaction-based includes CellPhoneDB, along with other approaches such CellChat[28] and IcellNet[29]. iTALK[47], singleCellSignalR[48] and NicheNet[49] also use curated interactions but, unlike CellPhoneDB, CellChat and IcellNet, they do not consider the structural composition of ligands and receptors in their analyses. To our knowledge, CellPhoneDB v5 and NeuronChat[50] are the only tools that account for the interactions mediated by non-peptidic ligands and GPCRs.

CellPhoneDB can prioritise interactions by leveraging information from spatial and chromatin accessibility assays. While tools like NicheNet[49] rely on TFs expression levels to infer cell interactions by weighting a gene regulatory network generated from general biological knowledge databases that are not curated, CellPhoneDB v5 adopts a different strategy. CellPhoneDB incorporates user-provided TF activities, which are typically derived from scATAC-seq data, and combines this with a manually curated collection of receptor-to-TFs relationships. The strength of this approach lies in its ability to determine when receptor activation, resulting from ligand binding, directly affects TF activation. Additionally, CellPhoneDB can leverage microenvironments defined with spatial techniques. These can restrict cell interaction inference to cells that colocalise. This contrasts with tools like Giotto[35], which infers cell interaction directly from undeconvoluted spatial spots.

Altogether, results from cell-cell communication inference methods should be employed cautiously and used to generate or support hypotheses. An initial biological question or hypothesis followed by downstream corroboration based on other *in silico* methods, complementary data modalities or experimental validations should be performed to obtain solid conclusions.

## MATERIALS

### Equipment

**Input data files for running CellPhoneDB**
- Expression file (mandatory). Normalised count data with gene identifiers either in Ensembl IDs, gene names or hgnc_symbol format. If subsampling is employed, users must indicate the data was log-transformed when using the subsampling option. For the scoring procedure, normalisation methods that transform zeros to any other value must be avoided. Formats accepted are .csv, .txt, .tsv, .tab and .pickle.

- Metadata file (mandatory) File with cell type annotation as defined by the user after. This file consists of two fields: 'Cell', indicating the name of the cell; and 'cell_type', indicating the name of the cluster considered. Formats accepted are .csv, .txt, .tsv, .tab and .pickle.

- Microenvironments file (optional). File with cells grouped per microenvironment. This file consists of two fields: 'cell_type', indicating the name of the cell type and 'microenvironment', indicating the microenvironment to which the cell belongs. Formats accepted are .tsv.

- Active Tfs file (optional). File with cell types and their active TF as defined by the user. This file consists of two fields: 'cell_type', indicating the name of the cell type and 'TF', indicating the TF active for the given cell. Formats accepted are .tsv.

- DEGs file (mandatory for method 3). File with cell types and differentially expressed genes as defined by the user. This file consists of two mandatory fields: 'cell_type', indicating the name of the cell type and 'gene', indicating the differentially expressed gene for the given cell. Formats accepted are .tsv.

**Hardware**
- Unix-based systems.

**Software**
- Python v.3.8 or higher (https://www.python.org/downloads/)
- Data reprocessing with user-prefered software, Seurat[51], SCANPY[52] or any other toolkits.

Examples of all the inputs required and installation can be found in the CellPhoneDB GitHub repository: https://github.com/ventolab/CellphoneDB

**Input data files for building the database.**
CellPhoneDB stores interaction and other properties of the interacting partners such as subunit architecture, TF-receptor interactions, and gene and protein identifiers in plain text format files. CellPhoneDB v5 maintains the same database structure used in previous versions[2] with the exception of two new files: transcription_factor_input.csv and uniprot_synonyms.tsv. There is no need for the user to modify these files, but it is important to note that these input files can be modified to add new user-provided interactions. CellPhoneDB v5 requires six mandatory files to generate the database:

- *gene_input.csv:* (as in previous versions) stores the relationships between the gene names, UniProt ids and ensembl ids. This information is retrieved from UniProt.
- *Interaction_input.csv:* (as in previous versions) stores the interactions and interaction-associated information such as the directionality, pathway classification and source of the interaction.
- *protein_input.csv:* (as in previous versions) stores information on the proteins participating in CellPhoneDB interactions. This information is retrieved from UniProt.
- *complex_input.csv:* (as in previous versions) stores the name given to the complexes and the set of proteins forming them; heterodimers or the non-protein ligands (last known enzyme + transporter).
- *transcription_factor_input.csv:* (new in v5) stores the relationships between receptors and their downstream TF. This is employed by the CellSign module to link the receptor to the user-provided TF activation status.
- *uniprot_synonyms.tsv:* (new in v5) stores the information to match gene name synonyms to the same protein name. The info has been retrieved from UniProt.

This version of the database only contains manually curated interactions and no longer collects general protein-protein interactions from third parties. To see the specific structure of the input files check CellPhoneDB v2[2]. Next, only new fields or input files are described.

The *gene_input.csv* keeps the same structure employed in the previous version of CellPhoneDB with the exception of five new non-mandatory columns.

- *version:* denotes the database version that incorporated the interaction.
- *interactors:* a string indicating in gene name format the partners participating in the interaction.
- *classification:* pathway classification for interactions. This classification is based on Reactome analysis following a manual reannotation. Only pathways grouping at least 2 interactions are kept.
- *directionality:* this column indicates the type and the directionality of the interaction order as 'partner A-partner B'. Possible fields are Ligand-Receptor, Ligand-Ligand, Receptor-Receptor, Adhesion-Adhesion and Gap-Gap.
- *modulatory_effect:* for interactions in which the effect of the interaction is of inhibitory characteristics, this is denoted with an 'Inhibitory' string.

The annotation described under the classification and directionality column is concatenated to the CellPhoneDB output files to facilitate interaction analysis.

The transcription_factor_input file contains manually curated information required to link the receptor to its *bona fide* TF. Mandatory fields in CellSign database (*transcription_factor_input.csv*) are: 'partner_receptor', 'partner_TF', 'protein_name_receptor', 'protein_name_TF', 'receptor_id' and 'TF_symbol'.

- *partner_receptor:* this field stores the receptor id which can have two formats: (i) for heteromeric proteins the complex name given in the *complex_input.csv* file is used while (ii) for homomeric receptors, the UniProt ID is used.
- *partner_TF:* UniProt id of the TF downstream the receptor.
- *protein_name_receptor:* UniProt protein name for the homodimeric receptors.
- *protein_name_TF:* UniProt protein name of the TF.
- *receptor_id:* receptor id, gene symbol for homodimeric proteins or name given in the receptor_input.csv for heterodimeric proteins..
- *TF_symbol:* gene symbol of the TF.
- *Effect:* effect of the receptor (i.e. activation or inhibition) on the TF activity.
- *Source:* literature source for the receptor-TF relationship.
- *Curator:* name of the receptor-TF relationship curator.
- *Xrefs_or_figures:* references supporting the receptor-TF relationship.
- *Brivanlou_class:* receptor-TF classification as in Brivanlou *et al.* 2022.

The uniprot_synonyms file stores and links gene names and their synonyms to UniProt identifiers. This input can be of relevance as some gene names change across the genome version.

- *Entry:* UniProt identifier.
- *Entry Name:* UniProt protein name.
- *Protein names:* full protein name.
- *Gene Name (synonyms):* synonym of the primary gene name if available.
- *Gene Names (primary):* primary gene name of the gene.

Output files

CellPhoneDB v5 stores results in plain text format files using the same structure as in previous versions of the tool. CellPhoneDB includes new output files to accommodate the new methodological implementations; method 3 (differential expression analysis), CellSign and scoring (Table 1).

- *relevant_interactions.txt*: this file denotes interactions for which at least one of the partners is differentially expressed and the other is expressed in more than k% of the cells.
- *CellSign_active_interactions.txt*: this file denotes which set of significant/relevant interactions have been found to have an active TF downstream of the receptor.
- *CellSign_active_interactions_deconvoluted.txt*: deconvoluted means expression value for each protein for those partners for which an active TF is found.
- *Interaction_scores.txt*: this file contains the results from the specificity scoring methodology developed for CellPhoneDB v5.

## PROCEDURE

**Analysis methods**

For an extensive description of the methods and examples on how to run CellPhoneDB v5 check the tutorials at https://github.com/ventolab/CellphoneDB.

**Installation. Timing 10 min**

CRITICAL: CellPhoneDB requires python 3.8 or higher.

1 Create conda environment and install CellPhoneDB

```
conda create -n cpdb python=3.8
conda activate cpdb
pip install cellphonedb
```

2 If CellPhoneDB is executed in Jupyter hubs, then add the environment to the kernel.

```
pip install -U ipykernel
python -m ipykernel install --user --name 'cpdb'
```

**Download database. Timing 1 min**

1 List database versions available

```
from IPython.display import HTML, display
from cellphonedb.utils import db_releases_utils

display(HTML(db_releases_utils.get_remote_database_versions_html()['db_releases_html_table']))
```

2 Download database

```
from cellphonedb.utils import db_utils
db_utils.download_database(cpdb_target_dir = 'path_db',
                           cpdb_version = 'v5.0.0')
```

**Update database. Timing 1 min**

CRITICAL: Database input files must follow the same format and directory structure as in the database provided by CellPhoneDB. The original input files can be used as templates for this.

**1 Create database from input files**

```
from cellphonedb.utils import db_utils
db_utils.create_db(cpdb_input_dir = 'dir_with_input_files')
```

**Run method 1. Timing 5 min**

**1 Define input files**

```
cpdb_file_path = 'db/v5/cellphonedb.zip'
meta_file_path = 'data/metadata.tsv'
counts_file_path = 'data/normalised_log_counts.h5ad'
microenvs_file_path = 'data/microenvironment.tsv'
out_path = 'results/cpdb_results'
```

**2 Running method 1 (simple analysis)**

```
from cellphonedb.src.core.methods import cpdb_analysis_method

cpdb_results = cpdb_analysis_method.call(
    cpdb_file_path = cpdb_file_path,
    meta_file_path = meta_file_path,
    counts_file_path = counts_file_path,
    counts_data = 'hgnc_symbol',
    output_path = out_path,
    separator = '|',
    threads = 5,
    threshold = 0.1,
    result_precision = 3,
    debug = False,
    output_suffix = None)
```

Optional arguments:

```
microenvs_file_path: [.tsv] file with microenvironment definition.
```

```
            score_interactions: [bool] indicates whether to score interactions.
```

**Run method 2 (statistical analysis). Timing 20-90 min**

**1 Define input files**

Same inputs files method 1 plus:
```
        active_tf_path = 'data/active_TFs.tsv'
```

**2 Running method 2**
```
        from cellphonedb.src.core.methods import
        cpdb_statistical_analysis_method

        cpdb_results = cpdb_statistical_analysis_method.call(
            cpdb_file_path = cpdb_file_path,
            meta_file_path = meta_file_path,
            counts_file_path = counts_file_path,
            counts_data = 'hgnc_symbol',
            iterations = 1000,
            threshold = 0.1,
            threads = 5,
            debug_seed = 42,
            result_precision = 3,
            pvalue = 0.05,
            separator = '|',
            debug = False,
            output_path = out_path,
            output_suffix = None)
```

Optional arguments:
```
      microenvs_file_path: [.tsv] file with microenvironment definition.
      score_interactions: [bool] indicates whether to score interactions.
      active_tfs_file_path: [.tsv] file with active TFs per cell.
      subsampling: [bool] defines whether to subsample cells with geometric
sketching.
      subsampling_log: [bool]: denotes if counts are log1p-trasnformed.
      subsampling_num_pc:[int]: number of PCs for subsampling.
      subsampling_num_cells: [int]: number of cells to subsample.
```

**Run method 3 (differential expression analysis). Timing 3 min.**

**1 Define input files**

Same input files as method 1 plus:
```
        degs_file_path = 'data/DEGs_genes.tsv'
```

```
        active_tf_path = 'data/active_TFs.tsv'
```

**2 Running method 3**
```
        from cellphonedb.src.core.methods import cpdb_degs_analysis_method

        cpdb_results = cpdb_degs_analysis_method.call(
            cpdb_file_path = cpdb_file_path,
            meta_file_path = meta_file_path,
            counts_file_path = counts_file_path,
            degs_file_path = degs_file_path,
            counts_data = 'hgnc_symbol',
            threshold = 0.1,
            result_precision = 3,
            threads = 4,
            separator = '|',
            debug = False,
            output_path = out_path,
            output_suffix = None)
```

Optional arguments:
```
        microenvs_file_path: [.tsv] file with microenvironment definition.
        score_interactions: [bool] indicates whether to score interactions.
        active_tfs_file_path: [.tsv] file with active TFs per cell.
```

**Run method 2 and 3 with scoring.**
CRITICAL: To properly score interactions, CellPhoneDB requires log-normalised expression data, any normalisation process (i.e. z-scaling) that transforms zeros to any other value must be avoided.
CRITICAL: It is possible to encounter interactions found significant but with scores as low as zero. This can occur due to multiple causes: (i) if one of the interacting genes is the least expressed in the dataset, during the scaling its value will be set to 0, (ii) the gene is expressed in less than the user-defined % of cells in the dataset and this parameter has not been taken into account for the differential expression analysis. Likewise, the scoring methodology might yield high scores for pairs of cells without relevancy. This can occur when none of the interacting partners are differentially expressed but their expression values are high.

**1 Define scoring argument**
```
        score_interactions = True
```

**2 Steps executed during scoring**

Step 1: Exclude genes not participating in any interaction and those expressed in less than k% of cells within a given cell type.

Step 2: Calculate the mean expression (G) of each gene (i) within each cell type (j).

$$G = \overline{X}_{ij}$$

Step 3: For heteromeric proteins, aggregate the mean gene expression of each subunit (n) employing the geometric mean.

$$H = \sqrt[n]{G_1 \cdot G_2 \ldots G_n}$$

Step 4: Scale means gene/heteromer expression across cell types between 0 and 10.

$$G_{scaled} = \frac{X - Xmin}{Xmax - Xmin} \cdot 10$$

Step 5: Calculate the product of the scale mean expression of the interaction proteins as a proxy of the interaction relevance.

$$Score = G_{ij,scaled} \cdot G_{ij,scaled}$$

TIMING

CellPhoneDB execution times vary depending on multiple factories such the size of the dataset, the methods of choice and the parameters selected. The use of microenvironments can drastically reduce running times. As a general approximation, for a 12GB dataset composed of 325,000 cells grouped in 42 different cell types and using 20 threads, the execution times for the different combinations of methods are:

Step 1: CellPhoneDB installation: ~10 minutes.
Step 2: Database download: ~1 minute.
Step 3: Running CellPhoneDB:
- Analysis with method 2 (statistical method):
  - Default parameters: ~90 minutes.
  - Subsampling to 10,000 cells and scoring: ~20 minutes.
  - Default parameters with scoring: ~95 minutes.
- Analysis with method 3 (DEG method):
  - Default parameters: ~90 seconds.
  - With scoring: ~200 seconds.

## TROUBLESHOOTING

Advice for common errors can be found in Table 2.

## Anticipated results

The new database and methods within CellPhoneDB v5 have been employed in various studies. First, to examine the interactions between epithelial and fibroblasts occurring in the human endometrium upon response to hormones[19]. Second, to examine how granulosa cells compartmentalise and drive oocyte differentiation[17]. Third, to investigate the interactions between invading trophoblast cells and maternal uterine cells during the early stages of human development. [18] Fourth, to identify interactions between trans-synaptic pacemaker cells and glial cells in the heart[16]. Fifth, to provide a platform to easily view and share cell-cell communication-related findings during the regeneration and differentiation of the human endometrium[20].

The updates of CellPhoneDB v5 facilitate the inference and prioritisation of cell-cell interactions by (i) leveraging data from other omics fields (e.g. spatial transcriptomics and chromatin accessibility); (ii) introducing new interaction modalities (e.g. GCPR and non-peptidic ligands); (iii) permitting more refined datasets queries (e.g. introduction of the differential expression method); (iv) rating interactions based on their specificity (e.g. scoring method); and (v) implementing new visualisation and sharing tools (e.g. ktplots and CellPhoneDBViz).

Taken together, the findings derived from CellPhoneDB v5 have been instrumental in producing new biologically relevant hypotheses. Some of these hypotheses have subsequently been experimentally validated.

**Code availability**

The CellPhoneDB code is available at [https://github.com/ventolab/CellphoneDB](https://github.com/ventolab/CellphoneDB).

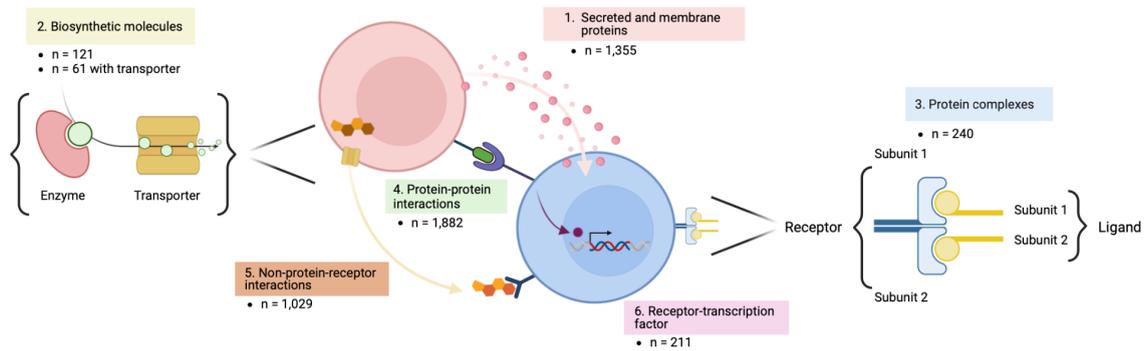

**Figure 1 | Overview of the database.** CellPhoneDB stores 1,419 interactors (i.e. proteins, enzymes and enzymes with transporter) of which (1) 1,056 are single-unit proteins, (2) 121 correspond to non-peptidic ligands and (3) 240 to heteromeric proteins. CellPhoneDB contains a total of 2,911 interactions of which (4) 1,882 are mediated through protein-protein interaction and (5) 1,029 are mediated by non-protein ligands. Finally, CellPhoneDB stores (6) 211 highly specific direct receptor-to-TF relationships. Created with BioRender.

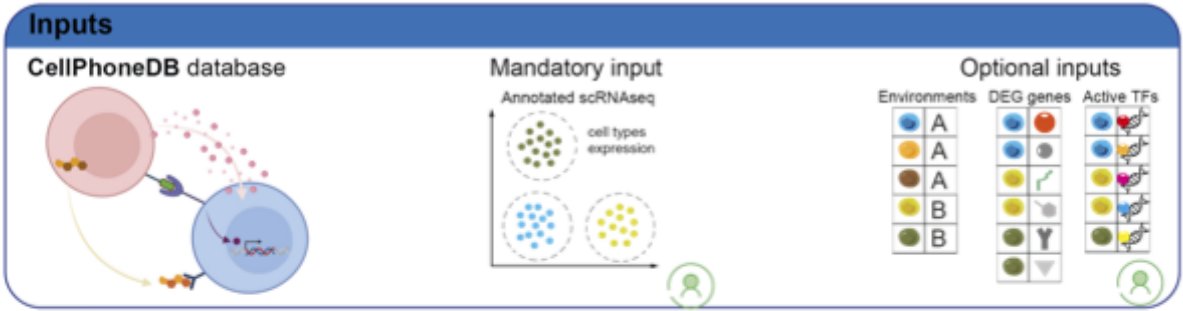
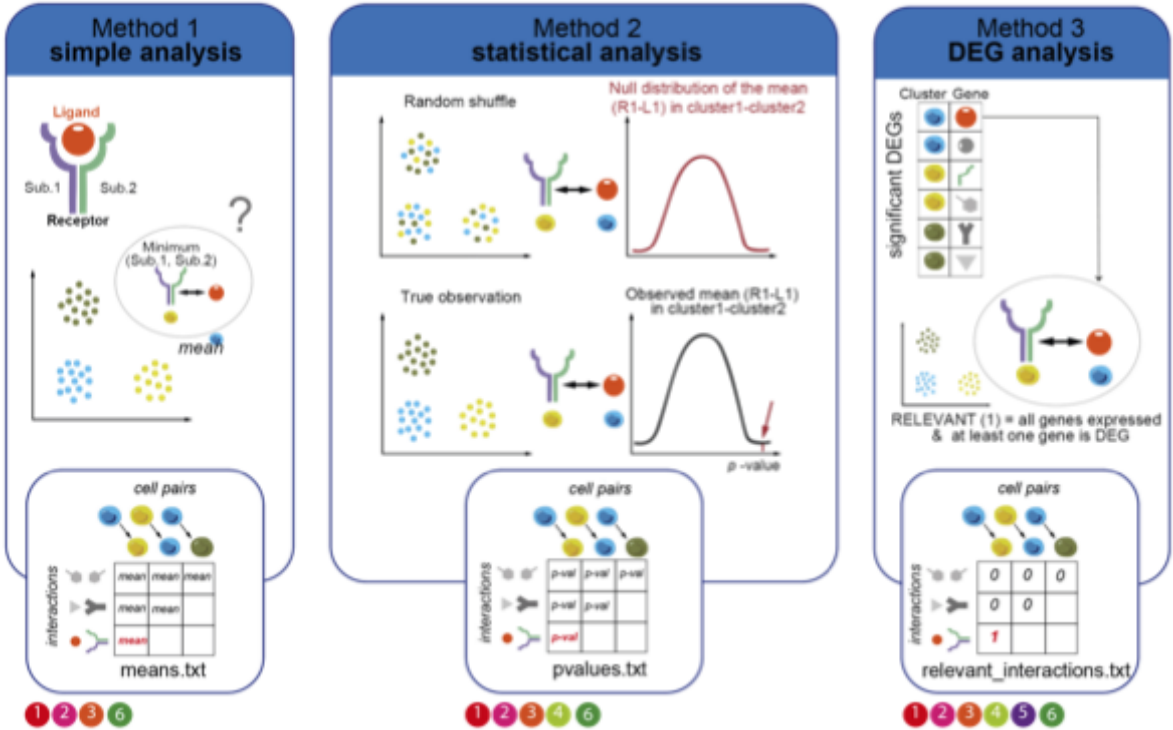
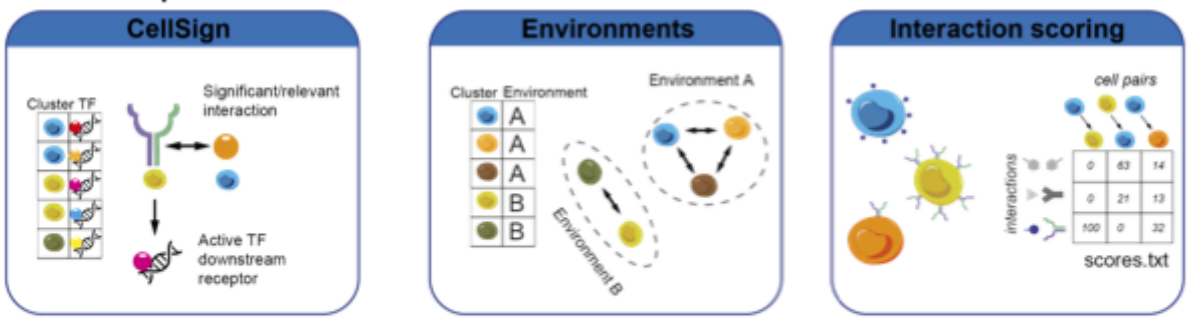
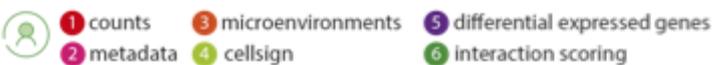

**Figure 2 | Overview of CellPhoneDB methods.** CellPhoneDB requires an annotated scRNAseq dataset and depending on the method, a set of input text files. CellPhoneDB has
22

three cell-cell interaction inference methods: (i) simple analysis that calculates the mean of the interacting partners, (ii) statistical analysis that tests the significance of the interacting partner's means by random shuffling, (iii) differential expression method that infers interactions based on a user-provided list of differentially expressed genes. CellPhoneDB incorporates three methods to prioritise the inferred interactions: (i) CellSign leverages interactions based on the status of the receptors downstream TF, (ii) environments narrow interactions to cells coexisting in space and time, (iii) scoring methodology ranks interactions based on the expression specificity of the interacting partners. Indicated by numeric indexes the inputs and prioritisation tools are compatible with each cell-cell interaction inference method.

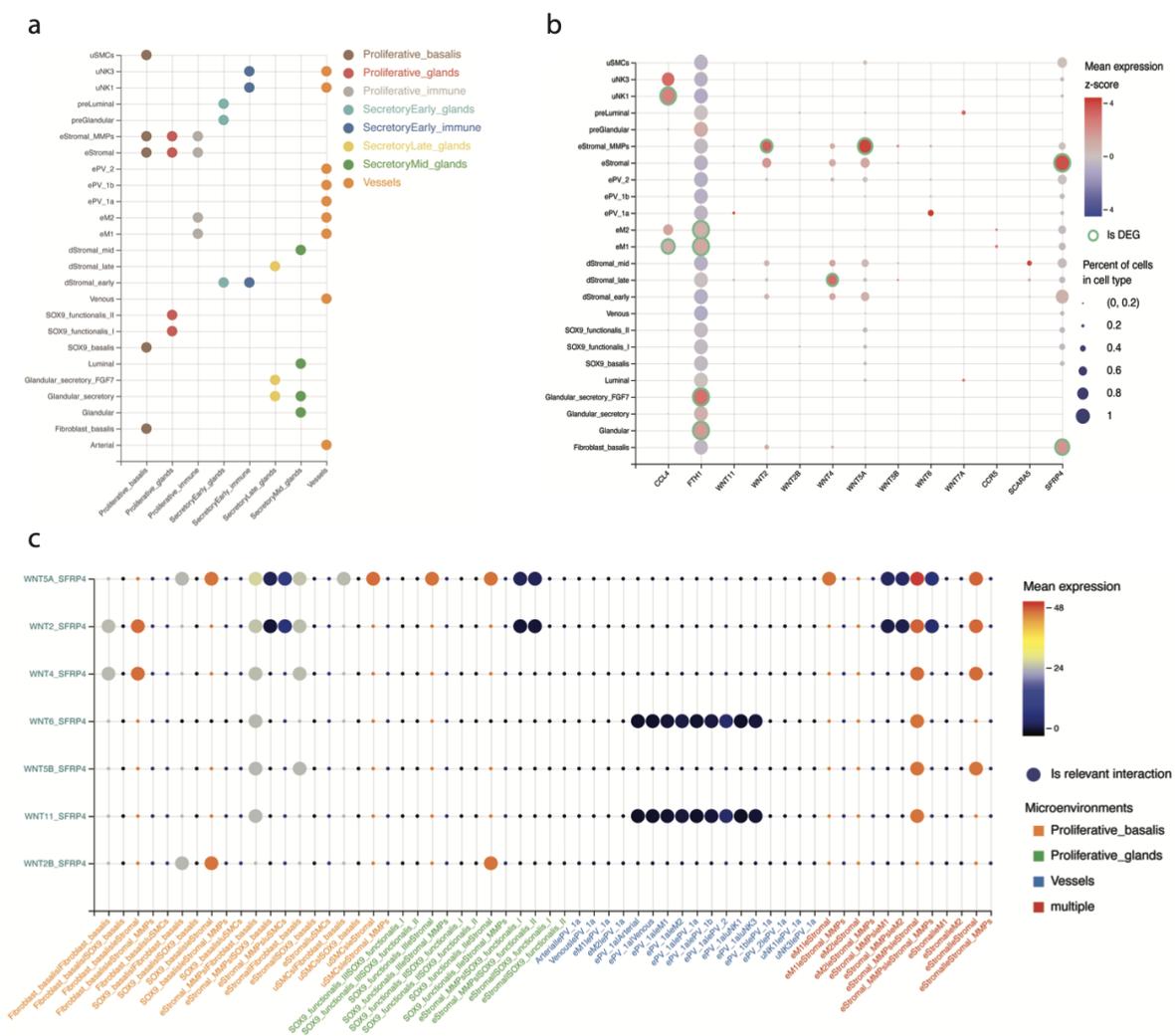

**Figure 3 | Overview of CellPhoneDBViz plots on the Harmonised endometrial atlas. a**, Dotplot depicting cell types (y-axis) per microenvironment (x-axis), some cell types can be part of multiple microenvironments. **b**, Z-scaled mean gene expression (x-axis) per cell type (y-axis). If the gene is differentially expressed (method 3) an outer green ring is shown. The

size of the dot is proportional to the percentage of cells of each cell type expressing the gene. **c**, Dot plot depicting the mean expression of the interacting partners (y-axis) per cell type pair (x-axis). Cell type pairs are coloured according to their microenvironment definition.

| Identifier | Definition | Output file | Example |
|---|---|---|---|
| id_cp_interaction | Unique CellPhoneDB identifier for each interaction stored in the database. | CellSign_active_interactions.txt; CellSign_active_interactions_deconvoluted.txt; interaction_scores.txt;relevant_interactions.txt; significant_interactions.txt | CPI-SS0A28DCA72 |
| interacting_pair | Name of the interacting pairs separated by '_' | CellSign_active_interactions.txt; CellSign_active_interactions_deconvoluted.txt; interaction_scores.txt;relevant_interactions.txt; significant_interactions.txt | CXCL12_CXCR4 |
| partner a or b | Identifier for the first interacting partner (A) or the second (B). It could be: UniProt (prefix 'simple:') or complex (prefix 'complex:'). | CellSign_active_interactions.txt; CellSign_active_interactions_deconvoluted.txt; interaction_scores.txt;relevant_interactions.txt; significant_interactions.txt | simple:P48061 |
| gene a or b | Gene identifier for the first interacting partner (A) or the.second (B). The identifier will depend on the input user list. | CellSign_active_interactions.txt; CellSign_active_interactions_deconvoluted.txt; interaction_scores.txt;relevant_interactions.txt; significant_interactions.txt | CXCL12 |
| secreted | True if one of the partners is secreted | CellSign_active_interactions.txt; interaction_scores.txt;relevant_interactions.txt; significant_interactions.txt | TRUE |
| receptor a or b | True if the first interacting partner (A) or the second (B) is annotated as a receptor in our database. | CellSign_active_interactions.txt; interaction_scores.txt;relevant_interactions.txt; significant_interactions.txt | FALSE |
| annotation_strategy | Curated if the interaction was annotated by the CellPhoneDB developers. Otherwise, the name of the database where the interaction has been downloaded from. | CellSign_active_interactions.txt; interaction_scores.txt;relevant_interactions.txt; significant_interactions.txt | curated |
| is_integrin | True if one of the partners is an integrin. | CellSign_active_interactions.txt; interaction_scores.txt;relevant_interactions.txt; significant_interactions.txt | FALSE |
| directionality | Directionality annotation as denoted in interaction_input file. | CellSign_active_interactions.txt; interaction_scores.txt;relevant_interactions.txt; significant_interactions.txt | Ligand-Receptor |
| classification | Pathway classification as denoted in interaction_input file. | CellSign_active_interactions.txt; interaction_scores.txt;relevant_interactions.txt; significant_interactions.txt | Signalling by Chemokines |
| active_TF | Denotes the active TF downstream the receptor. | CellSign_active_interactions_deconvoluted.txt | STAT1 |
| celltype_pairs | Denotes the pair of interacting cells for which the receptor-TF interaction is found active. | CellSign_active_interactions_deconvoluted.txt | PV MMP11|EVT_1 |
| active_celltype | Cell type expressing the active transcription factor. | CellSign_active_interactions_deconvoluted.txt | EVT_1 |
| active | Denotes whether downstream the receptor the transcription factor is active (1) or not (.) | CellSign_active_interactions.txt | 1 |
| score | Score generated by the scoring | interaction_scores.txt | 78.23 |

| | methodology. | | |
|---|---|---|---|
| relevancy | For method 3, denotes whether the interaction has been found relevant (1) or not (0) based on the differential expression of the interacting partners. | relevant_interactions.txt | 1 |
| P-value | For method 2, denotes the P-value obtained after random shuffling procedure. | significant_interactions.txt | 0.05 |

**Table 1 | Description of the fields present in the output files**: CellSign_active_interactions.txt, CellSign_active_interactions_deconvoluted.txt, Interaction_scores.txt relevant_interactions.txt and significant_interactions.txt

| Problem | Possible reason | Solution |
|---|---|---|
| KeyError: "['version', 'uniprot_5'] not in index" | Missing columns on one of the input files, possibly due to using a previous version of the database. | Add missing columns using as reference the original input files provided with CellPhoneDB v5. |
| [ERROR] Invalid Counts data | Order of the input count and metadata might be switched or genes are not in the correct format. | Add input data using the correct arguments and ensure gene Ids are in the specified format. |
| [Errno 13] Permission denied: 'results'. | User has no permissions to write on the output folder. | Add permissions to the folder to save results. |
| [SSL: CERTIFICATE_VERIFY _FAILED] | SSL certificate version doesn't match the required. | Upgrade to the latest SSL certificate. |
| Cell IDs in the meta file do not match those/exist in counts columns. | All cells in the meta file must be present in the count data. | Make sure the meta file is in the proper format and only contains cells present in the count matrix. |

**Table 2 | Troubleshooting table**.

**Author contribution statement**

K.T, L.G., S.T. and R.V.-T. conceived and developed the protocol and wrote the manuscript. M.P. and R.P. optimised CellPhoneDB v5 code, R.P. implemented, J.C. provided and curated the GPCRs CellPhoneDBViz, A.H and Z.K.T. developed ktplots and ktplotspy.

**ORCID for corresponding authors**

Roser (0000-0002-9870-8474)

Luz (0000-0002-7863-9619)

**Acknowledgements**

We thank J. Shilts for introducing new interactions and R. Vilarrasa for her feedback on the scoring methodology. We are grateful to the Vento-Tormo lab members for their fruitful advice. This project has been made possible in part by grant number 2022-249429 from the Chan Zuckerberg Initiative DAF, an advised fund of Silicon Valley Community Foundation, Wellcome Core (220540/Z/20/A), and UK Research and Innovation (UKRI) under the UK government's Horizon Europe funding guarantee EP/Y009924/1.

**Competing interests**

In the past 3 years, S.A.T. has consulted or been a member of scientific advisory boards at Roche, Genentech, Biogen, GlaxoSmithKline, Qiagen and ForeSite Labs, is an equity holder of Transition Bio, and is a co-founder of Ensocell Therapeutics.